\documentclass[12pt]{iopart}
\usepackage{times}
\usepackage{iopams}

\begin{document}

\title[Should recombinations of radical pairs be considered as accompanied by measurements?]{Should
recombinations of radical pairs be considered as accompanied by
measurements?}
\author{L V Il'ichov${}^{~a,c}$ and S V Anishchik${}^{~b,c}$}
\address{${}^{a~}$Institute of Automation and Electrometry SB RAS, 630090 Novosibirsk, Russia, \\
${}^{b~}$Institute of\, Chemical Kinetics and Combustion SB
RAS, 630090 Novosibirsk, Russia, \\
 ${}^{c~}$Novosibirsk State University, 630090 Novosibirsk, Russia.}
\ead{leonid@iae.nsk.su and svan@kinetics.nsc.ru}

\begin{abstract}
{The so-called ``phenomenological'' kinetic equation for one-pair
density operator for spin-selective reactions is defended. We
derive this equation from the kinetic equation for density
operator of \emph{all} pairs which are treated as singlet and
triplet bosons. There presented some reasons for inconsistency of
measurement-like approach to the problem.}
\end{abstract}

\section{Introduction}
We address the problem of correct treating and interpretation of
recombination events in the ensemble of radical pairs.
Particularly we deal with the relation between the spin-selective
recombinations of pairs and measurements of their spin.  Our main
statement concerns the radical distinction between a
measurement-like process and proper act of recombination. In
numerical works by Kominis (see \cite{Komin09} and references
therein) any recombination event is considered as that of
measurement-like. On the basis of this view-point his critics of
the well-known 'phenomenological' kinetic equation
\cite{Schulten2,Kniga}
\begin{equation}
\frac{d}{dt}\rho + \imath [\mathcal{H},\rho]= -k_{S}(Q_{S}\rho +
\rho\,{Q}_{S}) - k_{T}({Q}_{T}\rho + \rho\,{Q}_{T}) \label{1}
\end{equation}
is based ($\hbar$ = 1). Here $\rho$ is the density matrix of a
pair; $\mathcal{H}$ -- its Hamiltonian; $2k_S$ and $2k_T$ are the
rate constants of singlet and triplet recombinations respectively.
For simplicity we take into account only spin variables of the
pairs. In this case $\rho$ is an operator in a 4-dimensional
Hilbert space with the basis $\{|S\rangle, |T_{-}\rangle,
|T_{0}\rangle, |T_{+}\rangle\}$;
\begin{equation}
{Q}_{S} = |S\rangle\langle S|, \;\;{Q}_{T} = \sum_{\sigma = 0,\pm
1}|T_{\sigma}\rangle\langle T_{\sigma}| \label{2}
\end{equation}
are the projectors onto the corresponding spin subspaces. In
literature one can find a misleading interpretation of the last
two anticommutator terms in rhs of (\ref{1}) as remnants of a full
Lindblad-type structure \cite{Lind} upon elimination of the
so-called 'sandwich' gain term:
\begin{equation}
2Q\rho\, Q - {Q}\rho - \rho\,{Q}\;\; \mapsto\;\; - ({Q}\rho +
\rho\,{Q})
 \label{3}
\end{equation}
(here $Q$ stands for $Q_S$ as well as for $Q_T$). In accordance
with this interpretation the transformation (\ref{3}) is
considered as the elimination of the gain term because of
recombination. Note that the Lindblad-type structure in lhs of
(\ref{3}) provides strait associations with the
spin-measurement-like process. The central point of the Kominis'
approach is the restoration of the full Lindblad structures in
(\ref{1}) in the framework of the measurement paradigm.

We are going to show that there is a firm background behind the
equation (\ref{1}), which has practically nothing common with any
measurement-like process.

\section{Spin-selective recombinations}
Our starting point is the kinetic equation for the
\emph{multi-pair quantum state} $\hat{\varrho}$. In contrast with
$\rho$ this operates in the infinite-dimensional Hilbert space
with the Fock basis $\{|n_{S}, n_{T_{+}}, n_{T_{0}},
n_{T_{-}}\rangle\}$, where $n$ are the numbers of pairs in the
corresponding states. We should write the kinetic equation in
terms of annihilation, $\hat{a}_S$, $\hat{a}_{T_{\sigma}}$, and
creation, $\hat{a}^{\dag}_S$, $\hat{a}^{\dag}_{T_{\sigma}}$,
operators for singlet and triplet pair respectively. So we treat
any pair as an individual 'particle'. These 'particles' are
bosons, i.e. one has the following commutator relations:
\begin{equation}
[\hat{a}_S, \hat{a}^{\dag}_S] = 1;\;\;  [\hat{a}_{T_{\sigma 1}},
\hat{a}^{\dag}_{T_{\sigma 2}}] = \delta_{\sigma 1,\sigma 2};\;\;
[\hat{a}_{S}, \hat{a}^{\dag}_{T_{\sigma}}] = 0.
 \label{4a}
\end{equation}
The mathematical description of the process of spontaneous acts of
death of the 'particles' (the acts of recombinations) is known.
For the both types of recombinations (singlet and triplet ones) we
have the corresponding three-term Lindblad structure in rhs of the
kinetic equation:
\begin{equation}
\frac{d}{dt}\hat{\varrho}(t) + \imath [\hat{H},\hat{\varrho}(t)]=
k_{S}(2\hat{a}_S\hat{\varrho}(t)\,\hat{a}^{\dag}_S -
\hat{a}^{\dag}_S\hat{a}_S\hat{\varrho}(t) -
\hat{\varrho}(t)\,\hat{a}^{\dag}_S\hat{a}_S) +
 \label{4}
\end{equation}
$$
k_{T}\sum_{\sigma}(2\hat{a}_{T_{\sigma}}\hat{\varrho}(t)\,\hat{a}^{\dag}_{T_{\sigma}}
- \hat{a}^{\dag}_{T_{\sigma}}\hat{a}_{T_{\sigma}}\hat{\varrho}(t)
-
\hat{\varrho}(t)\,\hat{a}^{\dag}_{T_{\sigma}}\hat{a}_{T_{\sigma}})
$$
The Hamiltonian $\hat{H}$ in (\ref{4}) should not be confused with
$\mathcal{H}$ from (\ref{1}). The evaluation of this equation is
made in Appendix.

The one-pair density matrix $\rho$ from (\ref{1}) is obtained by
the following standard way:
\begin{equation}
\rho = |S\rangle\langle\hat{a}^{\dag}_S\hat{a}_S\rangle\langle S|
+
\sum_{\sigma}\bigg(|T_{\sigma}\rangle\langle\hat{a}^{\dag}_S\hat{a}_{T_\sigma}\rangle\langle
S| +
|S\rangle\langle\hat{a}^{\dag}_{T_\sigma}\hat{a}_S\rangle\langle
T_{\sigma}|\bigg) + \label{5}
\end{equation}
$$
\sum_{\sigma 1, \sigma 2}|T_{\sigma
1}\rangle\langle\hat{a}^{\dag}_{T_{\sigma 2}}\hat{a}_{T_{\sigma
1}}\rangle\langle T_{\sigma 2}|,
$$
where $\langle...\rangle = Tr(\hat{\varrho}(t)...)$. The mean
value of any multi-pair operator $\hat{O}$ evolves according to
the equation
\begin{equation}
\frac{d}{dt}\langle\hat{O}\rangle +
\imath\langle[\hat{O},\hat{H}]\rangle =
k_S\bigg(\langle\hat{a}^{\dag}_{S}[\hat{O},\hat{a}_{S}]\rangle +
\langle [\hat{a}^{\dag}_{S},\hat{O}]\hat{a}_{S}\rangle\bigg) +
\label{5b}
\end{equation}
$$
k_T\sum_{\sigma}\bigg(\langle\hat{a}^{\dag}_{T_{\sigma}}[\hat{O},\hat{a}_{T_{\sigma}}]\rangle
+ \langle
[\hat{a}^{\dag}_{T_{\sigma}},\hat{O}]\hat{a}_{T_{\sigma}}\rangle\bigg)
$$
Substituting the operator $\hat{O}$ with proper bilinear products
of $\hat{a}^{\dag}$ and $\hat{a}$, one gets the kinetic equation
for all elements of $\rho$:
\begin{equation}
\frac{d}{dt}\langle\hat{a}^{\dag}_S\hat{a}_S\rangle +
\imath\langle[\hat{a}^{\dag}_S\hat{a}_S,\hat{H}]\rangle =
-2k_S\langle\hat{a}^{\dag}_S\hat{a}_S\rangle \label{5a}
\end{equation}
\begin{equation}
\frac{d}{dt}\langle\hat{a}^{\dag}_S\hat{a}_{T_{\sigma}}\rangle +
\imath\langle[\hat{a}^{\dag}_S\hat{a}_{T_{\sigma}},\hat{H}]\rangle
= -(k_S + k_{T})\langle\hat{a}^{\dag}_S\hat{a}_{T_{\sigma}}\rangle
\label{5b}
\end{equation}
\begin{equation}
\frac{d}{dt}\langle\hat{a}^{\dag}_{T_{\sigma}}\hat{a}_S\rangle +
\imath\langle[\hat{a}^{\dag}_{T_{\sigma}}\hat{a}_S,\hat{H}]\rangle
= -(k_S + k_T)\langle\hat{a}^{\dag}_{T_{\sigma}}\hat{a}_S\rangle
\label{5c}
\end{equation}
\begin{equation}
\frac{d}{dt}\langle\hat{a}^{\dag}_{T_{\sigma 2}}\hat{a}_{T_{\sigma
1}}\rangle + \imath\langle[\hat{a}^{\dag}_{T_{\sigma
2}}\hat{a}_{T_{\sigma 1}},\hat{H}]\rangle =
-2k_T\langle\hat{a}^{\dag}_{T_{\sigma 2}}\hat{a}_{T_{\sigma
1}}\rangle \label{5c}
\end{equation}
Combining these equations with the use of (\ref{5}) and (\ref{2})
we arrive at an equation for $\rho$ with rhs identical to that of
(\ref{1}).

Its worth to stress that the equation (\ref{4}) describes the
process of spontaneous spin-selective 'annihilation' of pairs, not
measurement of their spins. 'If it were the case' picture of a
measurement-like process will be discussed later.

\section{Hypothetical measurement-like recombinations}

Let us consider an imaginary process in which recombination and
pair's spin measurement are really inseparable. It will be shown
that the evolution equation for the one-pair density matrix $\rho$
resembles neither (\ref{1}) nor the equation by Kominis
\cite{Komin09}.

Note that any measurement of a pair's spin create information
about the spin value. This information is written down in the
environment. Hence the pair appears to be entangled with the
environment. Let us first consider the following phenomenological
equation for $\rho$ where the pairs are subjected to spin
measurements but no recombinations take place:
\begin{equation}
\frac{d}{dt}\rho + \imath [\mathcal{H},\rho]= k(Q_{S}\rho\,Q_{S} +
Q_{T}\rho\,{Q}_{T} - \rho).  \label{100}
\end{equation}
Here $k$ is the number of measurement events per second. The first
two terms in the brackets in rhs of (\ref{100}) stand for the two
possible outcome of a measurement (singlet or triplet states). Any
singlet-triplet coherence vanish in accordance with rhs of
(\ref{1}). This is due to the mentioned entangling process. At the
same time the total number of pairs conserves: $Tr\rho = const$
(no recombinations). Now assume that upon a spin measurement with
the singlet outcome the pair recombines (in the singlet channel)
with the probability $p_S$, and the same for the triplet outcome.
One may account the loss of pairs due to these recombinations
eliminating parts of the gain terms in rhs of (\ref{100}):
\begin{equation}
\frac{d}{dt}\rho + \imath [\mathcal{H},\rho]=
k\bigg((1-p_S)Q_{S}\rho\,Q_{S} + (1-p_T)Q_{T}\rho\,{Q}_{T} -
\rho\bigg). \label{101}
\end{equation}
If we introduce the notations $\tilde{k}_S = p_Sk$, $\tilde{k}_T =
p_Tk$ and take into account that $Q_T = 1-Q_S$, we arrive at
\begin{equation}
\frac{d}{dt}\rho + \imath [\mathcal{H},\rho]= (2k - \tilde{k}_S -
\tilde{k}_T)Q_{S}\rho\,Q_{S} - (k - \tilde{k}_T)(Q_{S}\rho +
\rho\,{Q}_{S}) - \tilde{k}_T \rho. \label{102}
\end{equation}
The 'sandwich' as well as anticommutator terms are present. Note
that if $\tilde{k}_T = 0$ and $\tilde{k}_S = k$ the equation
(\ref{101}) lies just in between the same cases of (\ref{1}) and
equation (5) from \cite{Komin09}.

In the work \cite{Hore} the equation (\ref{102}) for the special
case $k = \tilde{k}_S + \tilde{k}_T$ was developed. The
comparative numerical analysis of the derived equation and that
one from (\ref{1}) revealed a slight difference.

\section{Conclusions}
Resuming we state that the kinetic equation for one-pair density
matrix $\rho$ in the case of spin-selective recombinations has
been shown to have the following form:
$$
\frac{d}{dt}\rho = -\imath [\mathcal{H},\rho] - k_{S}(Q_{S}\rho +
\rho\,{Q}_{S}) - k_{T}({Q}_{T}\rho + \rho\,{Q}_{T}).
$$
Our approach is only valid for geminate pairs. In the opposite
case (for possibility of cross-recombinations) one may not
consider the radical pairs as bosons.

We show that the kinetic equation for the measurement-like
recombination process differs radically both from the conventional
equation and from Kominis' results. We also do not share the
opinion from \cite{Hore} ``...that the the quantum measurement
approach should normally be used in future simulations of
spin-selective radical pair reactions''. From our view-point the
spin selective recombination is not a measurement. In the
hypothetic opposite case when any recombination is preceded by
spin measurement there emerges an information concerning the spin
of survived pairs. This options corresponds to the terms from
(\ref{101}) proportional to $1-p_S$ and $1-p_T$. There is no any
basis to suppose such a phenomenon takes place in recombination
process.

\section*{Appendix}
There is a regular derivation of equations like (\ref{4}) from the
first principles \cite{Glaub,CohTan}. As an example, we will give
a sketch of such a derivation for singlet Lindblad structure in
rhs of (\ref{4}). For example it can be obtained in the second
order of perturbation with respect to the following dynamics:
\begin{equation}
\hat{V}  = \lambda
\,\hat{a}_{S}\otimes(\mathbf{\hat{b}^{\dag}}\cdot
\mathbf{\hat{E}^{\dag}}) + H.c. \label{6}
\end{equation}
responsible for transformation of a singlet pair\,  into a product
$b$. This process is accompanied by photon emission.
Electromagnetic field is the vector operator
$\mathbf{\hat{E}^{\dag}}$ (this part is responsible for photon
creation). So the product should be vector as well, to make
$\hat{V}$ invariant under rotations, i.e. it should have total
unit angular momentum. Let $\hat{R}(t)$ be the total statistical
operators of pairs, products and photons. $\hat{R}(t)$ evolves
with respect to the following equation
\begin{equation}
\frac{d}{dt}\hat{R}(t) = -\imath[\hat{H}_{a} + \hat{H}_{b} +
\hat{H}_{ph} + \hat{V}, \hat{R}(t)], \label{7}
\end{equation}
where $\hat{H}_{a}$, $\hat{H}_{b}$ and $\hat{H}_{ph}$ are free
Hamiltonians of pairs, products and photons, respectively. In the
interaction representation
\begin{equation}
\hat{R}_{I}(t) = \exp[\imath(\hat{H}_{a} + \hat{H}_{b} +
\hat{H}_{ph})t]\hat{R}(t)\exp[-\imath(\hat{H}_{a} + \hat{H}_{b} +
\hat{H}_{ph})t], \label{8}
\end{equation}
\begin{equation}
\hat{V}_{I}(t) = \exp[\imath(\hat{H}_{a} + \hat{H}_{b} +
\hat{H}_{ph})t]\hat{V}\exp[-\imath(\hat{H}_{a} + \hat{H}_{b} +
\hat{H}_{ph})t] \label{9}
\end{equation}
(\ref{7}) turns into
\begin{equation}
\frac{d}{dt}\hat{R}_{I}(t) = -\imath[\hat{V}_{I}(t),
\hat{R}_{I}(t)]. \label{10}
\end{equation}
In the second order with respect to $\hat{V}_{I}(t)$ one has
\begin{equation}
\hat{R}_{I}(t + \Delta t) = \hat{R}_{I}(t) -\imath\int_{t}^{t +
\Delta t}dt'[\hat{V}_{I}(t'), \hat{R}_{I}(t)] - \label{11}
\end{equation}
$$
\int_{t}^{t + \Delta
t}dt'\int_{t}^{t'}dt''[\hat{V}_{I}(t'),[\hat{V}_{I}(t'')
\hat{R}_{I}(t)]]
$$
In the last term in rhs we replaced $\hat{R}_{I}(t'')$ with
$\hat{R}_{I}(t)$ due to the assumption $\Delta t \ll \tau_{rel}$,
where $\tau_{rel}$ is a typical relaxation time. There is another
time scale, $\tau_{corr}$ -- the correlation time which will
appear later. If $\Delta t \gg \tau_{corr}$ one can neglect all
correlations between pairs and their environment (products and
photons):
\begin{equation}
\hat{R}_{I}(t) = \hat{\varrho}_{I}(t)\otimes\hat{\varrho}_{b,ph}
\label{12}
\end{equation}
The statistical operator $\hat{\varrho}_{b,ph}$ is assumed
stationary. We also suppose that
\begin{equation}
\mathbf{\hat{E}}\hat{\varrho}_{b,ph} =
\hat{\varrho}_{b,ph}\mathbf{\hat{E}^{\dag}} = 0, \label{13}
\end{equation}
i.e. all emitted photons are rapidly absorbed so that
$\hat{\varrho}_{b,ph}$ is the vacuum with respect to
electromagnetic field.

From (\ref{11}) we get the 'coarse-grained' equation
\begin{equation}
\frac{\Delta\hat{\varrho}_{I}(t)}{\Delta t} = -\frac{1}{\Delta
t}\int_{t}^{t + \Delta
t}dt'\int_{t}^{t'}dt''Tr_{b,ph}[\hat{V}_{I}(t'),[\hat{V}_{I}(t''),
\hat{\varrho}_{I}(t)\otimes\hat{\varrho}_{b,ph}]].\label{14}
\end{equation}
Omitting details we arrive at
\begin{equation}
\frac{\Delta\hat{\varrho}_{I}(t)}{\Delta t} =
-\frac{|\lambda|^{2}}{\Delta t}\int_{0}^{\infty}d\tau\int_{t}^{t +
\Delta
t}dt'\bigg[g(\tau)\bigg(\hat{a}_{S}^{\dag}(t')\hat{a}_{S}(t'-\tau)\hat{\varrho}_{I}(t)
- \label{15}
\end{equation}
$$
\hat{a}_{S}(t'-\tau)\hat{\varrho}_{I}(t)\hat{a}_{S}^{\dag}(t')\bigg)
+
g(-\tau)\bigg(\hat{\varrho}_{I}(t)\hat{a}_{S}^{\dag}(t'-\tau)\hat{a}_{S}(t')
-
\hat{a}_{S}(t')\hat{\varrho}_{I}(t)\hat{a}_{S}^{\dag}(t'-\tau)\bigg)\bigg].
$$
Here
\begin{equation}
g(\tau) = g(-\tau)^{\ast} \doteq
Tr_{b,ph}(\mathbf{\hat{b}}_{I}(\tau)\cdot
\mathbf{\hat{E}}_{I}(\tau))(\mathbf{\hat{b}^{\dag}}_{I}(0)\cdot
\mathbf{\hat{E}^{\dag}}_{I}(0))\hat{\varrho}_{b,ph} \label{16}
\end{equation}
is the correlation function; $\tau_{corr}$ is its typical scale.
For simplicity we assume $\langle\hat{H}_{a}\rangle\tau_{corr} \ll
1$. Coming back to the Schrodinger representation we get
\begin{equation}
\frac{d}{dt}\hat{\varrho}(t) + \imath
[\hat{H}_{a},\hat{\varrho}(t)]=   \label{17}
\end{equation}
$$
(k_{S} +
\imath\kappa_{S})(\hat{a}_S\hat{\varrho}(t)\,\hat{a}^{\dag}_S -
\hat{a}^{\dag}_S\hat{a}_S\hat{\varrho}(t)) + (k_{S} -
\imath\kappa_{S})(\hat{a}_S\hat{\varrho}(t)\,\hat{a}^{\dag}_S -
\hat{\varrho}(t)\,\hat{a}^{\dag}_S\hat{a}_S),
$$
where
\begin{equation}
k_{S} = |\lambda|^{2}Re\int_{0}^{\infty}g(\tau)d\tau   \label{18}
\end{equation}
is the rate constant of singlet recombinations and
\begin{equation}
\kappa_{S} = |\lambda|^{2}Im\int_{0}^{\infty}g(\tau)d\tau
\label{19}
\end{equation}
is a parameter of a slight Hamiltonian renormalization:
\begin{equation}
\hat{H}_{a} \mapsto \hat{H}_{a} +
\kappa_{S}\hat{a}^{\dag}_S\hat{a}_S = \hat{H}. \label{20}
\end{equation}
We get the first line of (\ref{4}).

\end{document}